\documentclass[%
preprint,
 amsmath,amssymb,
 aps,
]{revtex4-2}

\usepackage{graphicx}
\usepackage{dcolumn}
\usepackage{bm}
\usepackage [english]{babel}
\usepackage [autostyle, english = american]{csquotes}

\begin{document}

This manuscript has been authored by UT-Battelle, LLC under Contract No. DE-AC05-00OR22725 with the U.S. Department of Energy. The United States Government retains and the publisher, by accepting the article for publication, acknowledges that the United States Government retains a non-exclusive, paid-up, irrevocable, world-wide license to publish or reproduce the published form of this manuscript, or allow others to do so, for United States Government purposes. The Department of Energy will provide public access to these results of federally sponsored research in accordance with the DOE Public Access Plan(http://energy.gov/downloads/doe-public-access-plan).

\title{Atomic dynamics in fluids: Normal mode analysis revisited}

\author{Jaeyun Moon}
\email{To whom correspondence should be addressed; E-mail: moonj@ornl.gov}
 \affiliation{Materials Science and Technology Division,\\ Oak Ridge National Laboratory, Oak Ridge, Tennessee 37831, USA}

\author{Lucas Lindsay}
 \affiliation{Materials Science and Technology Division,\\ Oak Ridge National Laboratory, Oak Ridge, Tennessee 37831, USA}
 

\author{Takeshi Egami}
\affiliation{Materials Science and Technology Division,\\ Oak Ridge National Laboratory, Oak Ridge, Tennessee 37831, USA\\
 Department of Materials Science and Engineering,\\ University of Tennessee, Knoxville, Tennessee 37996, USA\\
 Department of Physics and Astronomy,\\ University of Tennessee, Knoxville, Tennessee 37996, USA}

\date{\today}
\clearpage

\begin{abstract}

Developing microscopic understanding of the thermal properties of liquids is challenging due to their strong dynamic disorder, which prevents characterization of the atomic degrees of freedom. There have been significant research interests in the past few decades to extend the normal mode analysis for solids to instantaneous structures of liquids. However, the nature of normal modes that arise from these unstable structures is still elusive. In this work, we explore the instantaneous eigenmodes of dynamical matrices of various Lennard-Jones argon liquid/gas systems at high temperatures and show that the normal modes can be interpreted as an interpolation of $T \to \infty$ (gas) and $T = 0$ (solid) mode descriptions. We find that normal modes become increasingly collisional and translational, recovering atomistic gas-like behavior rather than vibrational with increase in temperature, suggesting that normal modes in liquids may be described by both solid-like and gas-like modes.

\end{abstract}

\maketitle
\clearpage

\section{Introduction}

In contrast to solids and gases, the nature of atomic dynamics of liquids remains elusive due to strong atomic correlations, leading to difficulties in microscopically understanding materials properties of liquid. Among other theoretical approaches that characterize atomic dynamics, normal mode analysis is widely used in various physical \cite{dove_introduction_1993} and life sciences \cite{case_normal_1994}. In normal mode analysis, one obtains normal mode/eigenmode frequencies and eigenvectors from diagonalization of dynamical matrices built from second derivatives of the interatomic potential of a given structure. Therefore, eigenmode frequencies represent local potential curvatures and eigenvectors depict constituent atomic motions relative to each other. 


For solids, normal modes are physically understood as harmonic oscillators about static equilibrium positions, with some exceptions including highly anharmonic systems \cite{sun_high_2020}. These harmonic oscillators are categorized as phonons, propagons, diffusons, and locons based on translational symmetry and mode interaction mechanisms. Normal mode analysis has been successful in describing various thermodynamic and thermal properties of solids, including free energy and heat capacity from a bottom-up perspective \cite{dove_introduction_1993, lindsay_survey_2018, mcgaughey_phonon_2019}. These properties can now be predicted with high accuracy from routine lattice dynamics calculations \cite{togo_first_2015}.

Encouraged by the success in describing vibrational properties of solids, many efforts have been made in recent years to extend the normal mode analysis to liquid systems where instantaneous structure snapshots are used instead of the lattice structure \cite{cho_instantaneous_1994, keyes_instantaneous_1997, seeley_isobaric_1991, seeley_normalmode_1989, gezelter_can_1997, clapa_localization_2012, schirmacher_modeling_2022, zaccone_universal_2021, stratt_instantaneous_1995}. These unstable snapshot structures (non-zero net forces) give rise to modes with imaginary frequencies in addition to modes with real frequencies. As such, a lot of research has been conducted to understand the nature of imaginary modes and various aspects of these have been used to describe some non-equilibrium processes including atomic diffusion \cite{seeley_isobaric_1991, zhang_what_2019}, glass transition \cite{bembenek_instantaneous_1995, bembenek_instantaneous_2001}, and melting \cite{melzer_instantaneous_2012}, with varying degrees of success. Despite these efforts, physical understanding of imaginary modes and their relation to various materials properties remains elusive. Real modes in liquids are often interpreted as extrapolation of real modes in solids as harmonic oscillators with well-defined frequencies in a local potential well. However, this assumption becomes particularly questionable at high temperatures as the structures themselves change much faster than most real mode periods. Accurate understanding of the nature of real modes in liquids is, therefore, also lacking.

To build a better understanding of normal mode behaviors, particularly for non-static structures (i.e., liquids and gases), we study normal mode characteristics of molecular dynamics (MD) snapshots of various argon systems from solid to gas at extremely high temperatures, up to $10^8$ K, at constant volume. In particular, we explore the high temperature limit ($T \to \infty$) of normal modes such that physical interpretations of normal modes at finite, intermediate temperatures can potentially become an interpolation between solid and gas rather than an extrapolation from solid behavior alone. We find that as temperature increases, normal modes involving diatomic units become a dominant feature in the normal mode spectra. Many of these pairs of atoms have positive potential energies and large kinetic energies such that normal modes with eigenvectors that are parallel with the `bond' can be described as collisional rather than vibrational. Further, we also observe modes progressively being projected to a single atom, mimicking single atom translating in space. Our findings, therefore, set the scale of the spectrum of eigenmodes of dynamical matrices where in one end, we have solid-like vibrational normal modes and in the other end we have gas-like collisional and translational modes.

\section{Methods}

Crystalline argon of 1372 atoms in the FCC structure and with a lattice parameter of 5.27 \AA \ was used as the initial input structure for MD simulations using LAMMPS \cite{plimpton_fast_1995} to generate equilibrated atomic structures at various temperatures at constant volume. Interatomic interactions are described by the Lennard-Jones potential \cite{jones_determination_1924, jones_determination_1924-1, rahman_molecular_1976}. The parameters used were $\epsilon$ = 0.0104 eV and $\sigma$ = 3.4 \AA. The cutoff distance was 2.5$\sigma$ and the potential was shifted to eliminate discontinuity. Timesteps of 0.1 fs (up to 10\textsuperscript{6} K) and 0.01 fs (10\textsuperscript{7} K and above) were used. At each temperature, all systems except that of the glass system were equilibrated for $5 \times 10^6$ timesteps in the canonical ensemble (NVT) prior to data recording of another $5 \times 10^6$ timesteps under the same ensemble. The equilibrated atomic structures (snapshots) generated from MD were used to perform lattice dynamics calculations using GULP \cite{gale_gulp:_1997} to obtain eigenfrequencies and eigenvectors of the $\Gamma$ point dynamical matrices with the entire domain considered as a unit cell. For better statistics, ten such structures at each temperature from MD simulations were used for the lattice dynamics calculations. 

Temperatures considered were 1 K for crystal and glass systems and ranging from $10^3$ to $10^8$ K in increments of factors of ten for high temperature liquids and gases. In this work, we consider all systems at 10\textsuperscript{6} K and above as gas states as their specific heat is $\frac{3}{2}Nk_B$ within numerical errors where $N$ is the number of atoms. Even though the physical density is kept constant, the effective atomic radius, where the potential energy equals $k_BT$, thus the effective packing fraction, becomes much reduced at high temperatures, as in gas. Thus, for simplicity, we loosely refer to all other high temperature systems below 10\textsuperscript{6} K as liquids. The glass system was created by a standard melt-quench technique (1000 K melting followed by an instantaneous quench to 1 K). This fast cooling rate was required due to typical poor glass formability of monatomic systems. Quenched glass was then equilibrated for $2 \times 10^7$ timesteps before data recording for  $5 \times 10^6$ timesteps.

\begin{figure}[h!]
 	\centering
 	\includegraphics[width=0.7\linewidth]{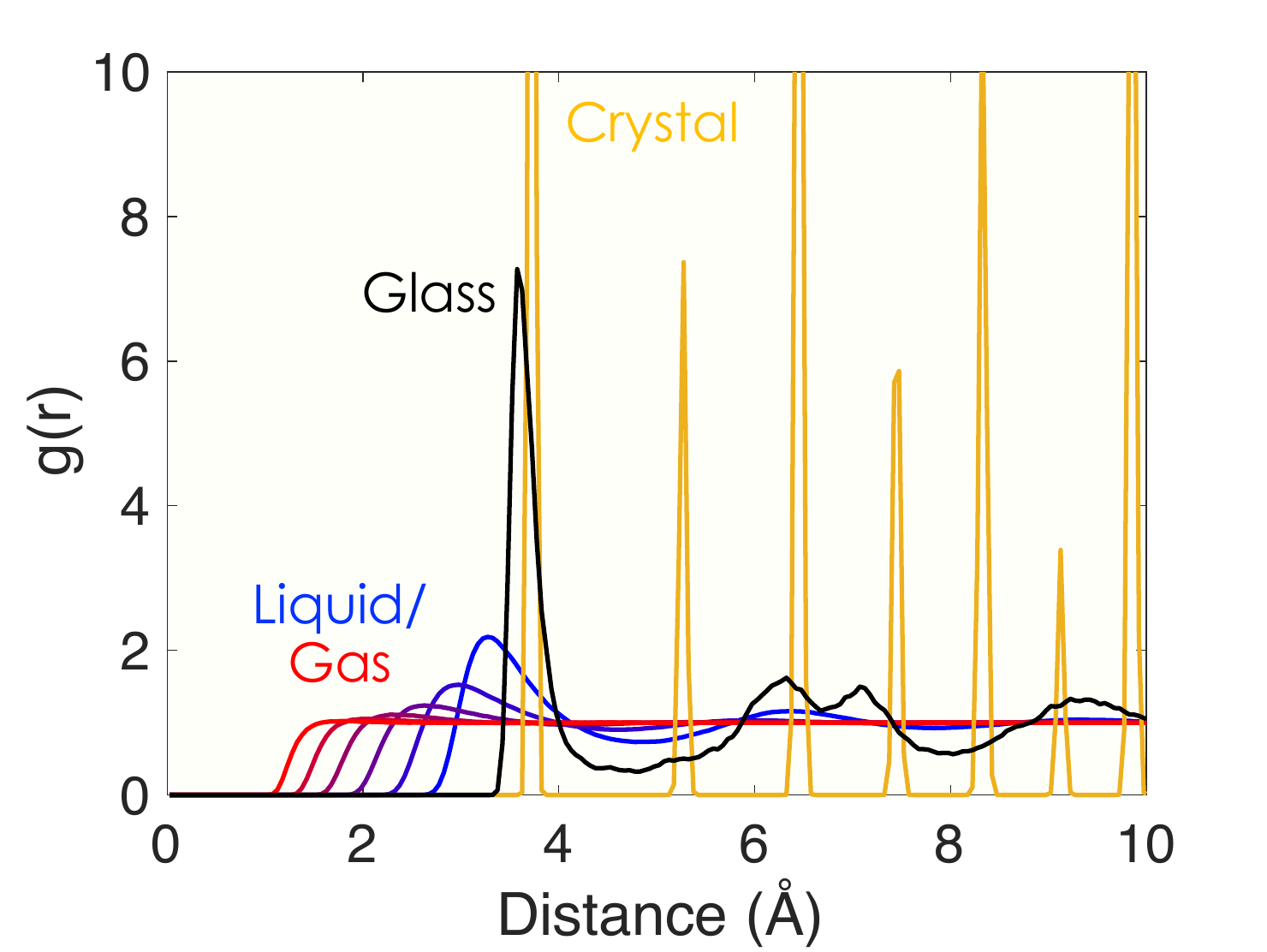}
 	\caption{Pair distribution functions of various argon phases from 1 K to 10\textsuperscript{8} K under constant volume. 1 K crystal (yellow curve) has very sharp peaks, some exceeding $g(r) = 30$ (not shown). Broadened peaks and valleys are shown for 1 K glass (black curve) demonstrating short and medium range order. From liquid to gas, progressive disappearance of peaks and valleys are observed with increase in temperature. Temperature ranges are from 10\textsuperscript{3} (blue curve) to 10\textsuperscript{8} K (red curve) with increments of a factor of 10.}
 	\label{fig:rdf_crystal_glass_liquid_gas}
 \end{figure}

To characterize the phases of the various structures, the pair distribution function for each system (shown in Fig. \ref{fig:rdf_crystal_glass_liquid_gas}) was calculated by 
\begin{equation}
    g(r) = \frac{1}{4\pi N n r^2}\sum_{i,j} \langle \delta(r - |\boldsymbol{r}_i - \boldsymbol{r}_j|)\rangle
\end{equation}
where $N$, $n$, and $\boldsymbol{r}_i$ are number of atoms, number density, and the atomic position of atom $i$, respectively. Sharp peaks are observed for the 1 K crystal and broadened peaks and valleys are observed for the 1 K glass demonstrating short and medium range orders are still present in the structure. For the liquid systems, peaks and valleys become broader with increase in temperature and eventually become step-functions similar to hard-sphere or dilute gases observed in the literature \cite{mcquarrie_statistical_2000}. The effective diameters of argon atoms decrease with increase in temperature as high kinetic energy enables atoms to approach close to each other, overcoming more of the repulsive portion of the potential, thus mimicking a dilute gas. 

From the snapshot structures of all systems, we perform lattice dynamics calculations to obtain eigenmode frequencies ($\omega_n$) and eigenvectors ($\boldsymbol{e}_n$) from $\Gamma$  point dynamical matrices ($\boldsymbol{D}$) as
\begin{equation}
    \omega^2_n \cdot \boldsymbol{e}_n = \boldsymbol{D} \cdot \boldsymbol{e}_n
\end{equation}
where subscript $n$ denotes a particular mode from 1 to $3N$ ($N = 1372$ for our systems). Dynamical matrices are related to the potential ($U$) by 
\begin{equation}
    D_{\alpha \beta, ij} = \frac{1}{\sqrt{m_i m_j}} \frac{\partial^2 U_{ij}}{\partial u_{\alpha,i} \partial u_{\beta,j}}
    \label{Eq: DM}
\end{equation}
where $m_i$ is the mass of atom $i$ and Greek subscripts denote Cartesian directions. There are no phase factors in Eq. \ref{Eq: DM} as the wavevector under consideration is zero and there are no lattice vectors for our non-periodic systems. All eigenvectors are real and are orthonormal such that $\boldsymbol{e}_n^T \cdot \boldsymbol{e}_{n'} = \delta_{n,n'}$. As dynamical matrices are Hermitian, all eigenvalues ($\omega_n^2$) are real, while eigenfrequencies ($\omega_n$) can be real or imaginary. We denote imaginary frequencies as negative values in figures below. Overall distributions of available eigenmode populations known as the density of states $g(\omega)$,  can then be found by 
\begin{equation}
    g(\omega) = \sum_{n = 1}^{3N} \delta(\omega - \omega_n)
\end{equation}

Using the mode eigenvectors, we can further characterize each mode by its inverse participation ratio (IPR) and phase quotient (PQ). IPRs ($p_n^{-1}$) are found by \cite{bell_atomic_1970}
\begin{equation}
    p_n^{-1} = \sum_i \Big( \sum_\alpha e_{i\alpha, n}^* e_{i\alpha, n} \Big)^2
    \label{Eq: IPR}
\end{equation}
where $e_{i\alpha, n}$ is the eigenvector component for atom $i$ in direction $\alpha$ for the mode $n$.  IPR gives a measure of how many atoms participate appreciably in the motion of a particular mode; therefore, IPR has been used to qualitatively provide information about the degree of localization of modes for various complex systems \cite{moon_sub-amorphous_2016, deangelis_thermal_2018} and even scattering behaviors of acoustic and optic phonons in binary compound crystals \cite{lindsay_phonon-isotope_2013}. The IPR is conventionally interpreted such that $1/N$ describes all atoms participating in the mode (delocalized) and 1 if the mode motion is completely localized to a single atom.

Phase quotients have also been useful in characterizing normal modes in various materials, especially disordered solids \cite{bell_acoustic_1975, allen_diffusons_1999, seyf_importance_2018} and is defined as \cite{bell_acoustic_1975}
\begin{equation}
    PQ_n = \frac{\sum_{i,j \in m} \sum_\alpha e_{i\alpha, n} e_{j\alpha, n}}{\sum_{i,j \in m} \Big|\sum_\alpha e_{i\alpha, n} e_{j\alpha, n}\Big|}
\end{equation}
where the sum over $m$ represents counting only the nearest neighbor pairs among atoms $i$ and $j$. PQ is a measure of the phase relationship of an atom and its neighbors for a given mode. A PQ value of 1 then represents cluster of atoms all moving in-phase, which is the case for Goldstone translational modes, and a PQ value of -1 means that all atoms participating in the mode are moving in the opposite phase/direction with their nearest neighbors. For our systems, we define nearest neighbors to be within $d_{c} = 2r_c$ where $\frac{4}{3}\pi r_c^3 = \frac{V}{N}$; however, we find that our results are not sensitive to the choice of $d_c$.

\section{Results}

\subsection{Normal modes in solids}

\begin{figure}[h!]
 	\centering
 	\includegraphics[width=0.7\linewidth]{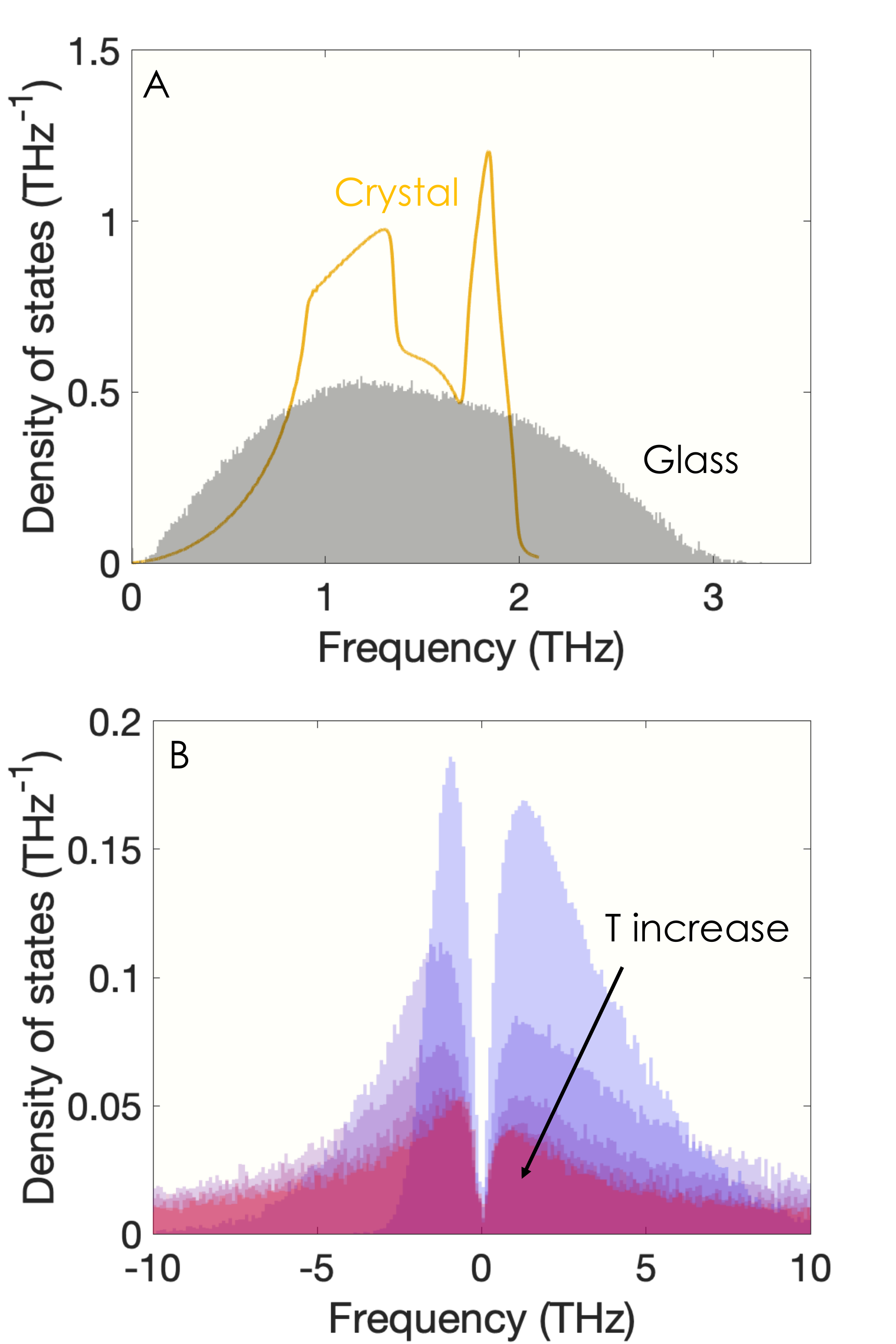}
 	\caption{Densities of states of normal modes for (A) 1 K solids and (B) high temperature liquids and gases from 10\textsuperscript{3} K (blue shade) to 10\textsuperscript{8} K (red shade) with increments of a factor of 10. Negative frequencies denote imaginary modes. We observe more dominant real modes at low temperatures and increasing importance of imaginary mode populations as measured by the area with increase in temperature. Density of states for glass argon is broadened (black shade) and has a wider frequency range compared to that of crystal argon (orange curve). }
 	\label{fig:DOS}
 \end{figure}

\textit{Density of states}. -- Eigenmode densities of states (DOS) for all systems are shown in Fig. \ref{fig:DOS}. Modes with real and imaginary frequencies are plotted on the positive and negative sides of the horizontal axis, respectively. For solids, all modes have real frequencies as shown in Fig. \ref{fig:DOS}A. Crystalline DOS computed from integration of wavevector points using the primitive unit cell (one atom per unit cell with primitive lattice vectors: $a_1 = (0.5, 0.5, 0), a_2 = (0, 0.5, 0.5),$ and $a_3 = (0.5, 0, 0.5)$) are also shown as reference. Sharp kinks in the densities of states for crystals are referred to as van Hove singularities where $\frac{d\omega}{dk} = 0$, mostly near zone-boundaries \cite{kittel_introduction_1976}. Crystalline argon has three acoustic branches with frequencies ranging up to 2 THz. There are no optical branches due to the primitive unit cell having only one atom. For the rest of the work, we utilize supercell argon crystal normal modes for analysis for more appropriate comparisons with the other non-periodic systems of the same size. For the glass system, there are no clear Brilliouin zone-boundaries nor translational symmetry; thus, it has a much broader density of states, though still has three Goldstone modes with zero frequency. The frequency ranges up to 3 THz, larger than that of crystalline argon likely due to some bonds being shorter than the mean equilibrium bond lengths.

\textit{Inverse participation ratio}. -- Crystalline modes are expected to be spatially delocalized whereas glasses are expected to have some localized modes due to structural disorder. Inverse participation ratios for both are shown in Fig. \ref{fig:IPR}A and give a measure of the degree of localization for each mode. In the literature, a normal mode has been considered to be localized when less than 10 to 20\% of atoms are participating in the mode motion \cite{allen_diffusons_1999, seyf_method_2016, aryana_tuning_2021}. For our purposes here, modes with IPR values over IPR\textsubscript{loc} = 0.007 (corresponding to 10\% atom participation) are considered to be localized. At 0 THz, we have the Goldstone modes with IPR = $1/N$ for both solids where all atoms are participating in the purely translational motion. As expected, all modes for crystalline argon have IPR values less than IPR\textsubscript{loc}, demonstrating that all modes are delocalized. For our glass system, most low frequency modes are also delocalized with some exceptions known as quasi-localized or resonant modes. Prior works have proposed that some localization at low frequencies is due to finite size effects \cite{feldman_numerical_1999}. At high frequencies, we observe a significant number of modes with IPR values greater than IPR\textsubscript{loc}, similar to that observed previously in another single element glass (amorphous silicon) \cite{allen_diffusons_1999}. These localized vibrational modes, often referred to as locons in the glass literature, form a small fraction of the total number of modes, about 2 to 3\%. Locons typically have low thermal diffusivities and conductivities compared to extended modes \cite{allen_diffusons_1999, moon_propagating_2018, kim_origin_2021}. Maximum IPR values found in our glass system are around 0.3.  

\begin{figure}[h!]
 	\centering
 	\includegraphics[width=1\linewidth]{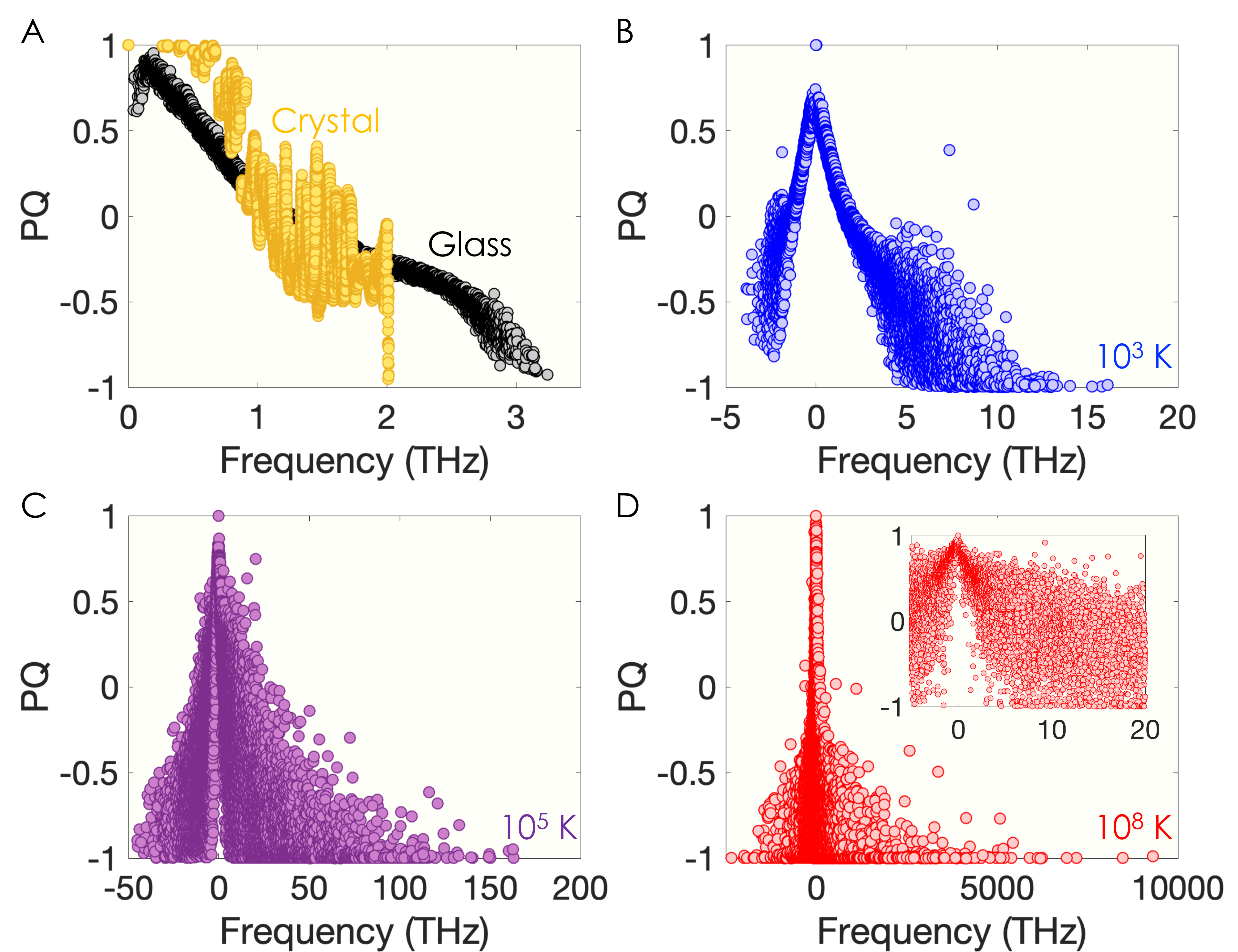}
 	\caption{Phase quotients (measure of motion of atoms compared to neighbors) for (A) 1 K solids (yellow circles for crystal and black circles for glass) and (B-D) select high temperature liquids and gases at 10\textsuperscript{3} K (blue circles), 10\textsuperscript{5} K (purple circles), and  10\textsuperscript{8} K (red circles). PQ = 1 corresponds to in-phase, while PQ = -1 corresponds to out-of-phase. Negative frequency denotes imaginary modes. Inset plot in (D) x-axis is scaled the same as (B) for comparisons.}
 	\label{fig:PQ}
 \end{figure}

\textit{Phase quotient}. -- Modal phase quotients (PQ) for our solid argon systems that characterize relative mode motion of an atom with its neighbors are shown in Fig. \ref{fig:PQ}A (PQ=1 for fully in-phase and PQ=-1 for fully out-of-phase). For the crystal system, some modes, including the 0 THz Goldstone modes, have PQ $\sim$ 1 at low frequencies demonstrating that atoms are moving nearly in phase with their neighbors. With increase in frequency, PQ decreases and eventually appears to fluctuate between PQ = 0 and PQ = -0.5 and approaches PQ $\sim$ -1. While crystalline argon does not have optical modes (typical out-of-phase vibrations in compound crystals), away from 0 THz acoustic modes develop phase differences among neighboring atoms which underlies the decrease in PQ values. For the argon glass, we observe a monotonic decrease in PQ values with increase in frequency away from 0 THz Goldstone modes that have PQ = 1, similar to other glasses reported in the literature \cite{allen_diffusons_1999, seyf_importance_2018}. The lowest PQ values found in the glass system are around -0.9, close to the fully out-of-phase value of PQ = -1. Prior works utilizing PQs in glasses have proposed that modes with PQ $>$ 0 and PQ $<$ 0 represent acoustic-like and optical-like modes, respectively \cite{allen_diffusons_1999, seyf_importance_2018}. Using this notation, an acoustic to optical like mode transition occurs around 1.2 THz for argon glass. 

With this normal mode analysis for solids as background context, we next characterize normal modes in the same way for various liquid and gas systems at higher temperatures.

\subsection{Normal modes in liquids and gases}

\textit{Density of states}. -- Normal mode densities of states for all liquid and gas systems are shown in Fig. \ref{fig:DOS}B. Similar to the solids, densities of states are scaled such that the area under the curve is 1. Only the low frequency region $|\omega|$ $<$ 10 THz is shown for ease of comparisons. Perhaps, the most notable difference in the normal mode densities of states of liquids and gases compared to those of solids is the significant presence of imaginary modes. We see a clear temperature dependence: the imaginary part of the spectra becomes more prominent with increase in temperature as measured by the area. However, as noted previously \cite{moon_microscopic_2022}, the rate at which this happens slows down at high temperatures as demonstrated in Fig. \ref{fig:DOS}B. We have previously proposed that equal number of imaginary modes and real modes marks a transition to a gas-like phase \cite{moon_microscopic_2022}. With increase in temperature, also the frequency ranges increase significantly up to 10\textsuperscript{4} THz and down to -2500 THz ($i\omega$) for the 10\textsuperscript{8} K gas but with very small probability.

Following Zwanzig's approach to relate self-diffusion coefficients to normal modes \cite{zwanzig_relation_1983}, some prior works have tried to express self-diffusion coefficient with the fractions of imaginary modes and real modes over all modes \cite{seeley_normalmode_1989, seeley_isobaric_1991, nave_instantaneous_2000, melzer_instantaneous_2012, clapa_localization_2012}. Some other works have tried to relate normal modes in non-solids to their specific heats \cite{baggioli_explaining_2021, moon_microscopic_2022} and glass transitions \cite{bembenek_instantaneous_1995, bembenek_instantaneous_2001}. Extrapolating the physical interpretation of real modes in solids to non-solids, it is often assumed that for real modes, liquid atoms vibrate as harmonic oscillators within a local potential well determined from the cage of nearest neighbor atoms \cite{melzer_instantaneous_2012, baggioli_explaining_2021, zaccone_universal_2021}.  However, physical interpretation of both real and imaginary modes in liquids is not clear and motivated our study. 

We aim below to characterize the high temperature limit of eigenmodes of dynamical matrices ($T \to \infty$) under the same density as those of solids such that interpretation of eigenmodes of liquids may potentially become an interpolation from both $T = 0$ and $T \to \infty$ eigenmodes rather than an extrapolation only from the $T = 0$ crystalline modes that we are familiar with.  

\begin{figure}[h!]
 	\centering
 	\includegraphics[width=1\linewidth]{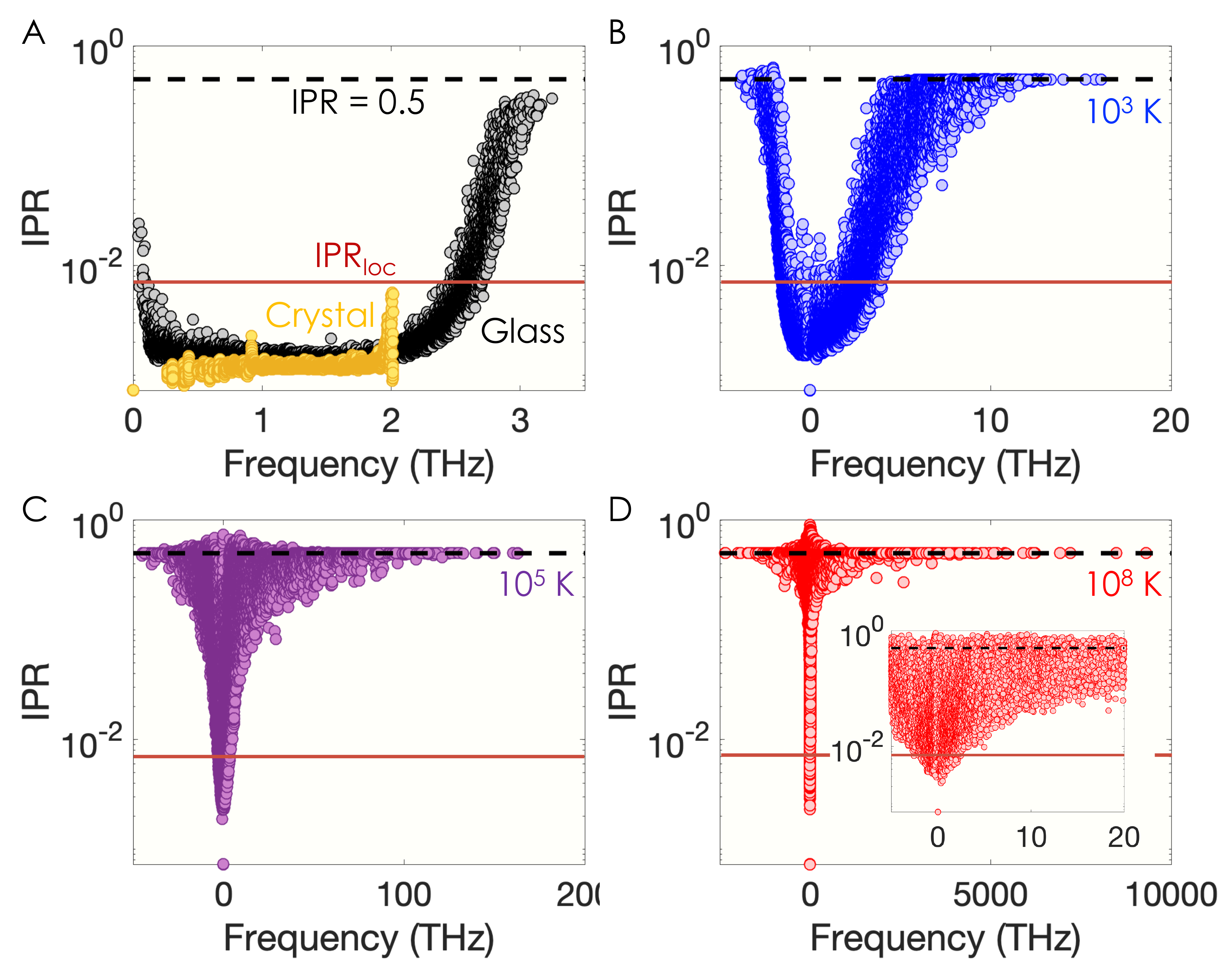}
 	\caption{Inverse participation ratios (measure of localization of modes) for (A) 1 K solids (yellow circles for crystal and black circles for glass) and (B-D) select high temperature liquids and gases at 10\textsuperscript{3} K (blue circles), 10\textsuperscript{5} K (purple circles), and  10\textsuperscript{8} K (red circles). IPR = $1/N$ corresponds to fully delocalized (all atoms involved), IPR = 1 corresponds to localized to a single atom. Negative frequencies denote imaginary modes. Dashed black line is at IPR = 0.5 (two-body modes) and solid brown line represent IPR = IPR\textsubscript{loc}. Modes with IPR $>$ IPR\textsubscript{loc} are considered to be localized. Inset plot in (D) x-axis is scaled the same as (B) for comparisons.}
 	\label{fig:IPR}
 \end{figure}

\textit{Inverse participation ratio}. -- Similar to solids, we characterize eigenmodes of liquids and gases via inverse participation ratios and phase quotients. Inverse participation ratio spectra at select high temperatures ($10^3$ K, $10^5$ K, and $10^8$ K) are shown in Fig. \ref{fig:IPR}B-D. At $10^3$ K, we see an asymmetric U shape with high real and imaginary frequency modes being strongly localized and low frequency modes being more extended. This IPR behavior mimics the glass spectral features, although the liquid system has a large component of imaginary modes. Similar trends have been previously shown in other liquid systems \cite{zhang_what_2019}. In the limit of high frequency, we seem to approach IPR = 0.5. With increase in temperature, low frequency modes have higher IPR values, filling up the empty region in the U shapes found at low temperatures. For our highest temperature system, we have a cross-like shape with the vertical limits dictated by the lowest IPR ($1/N$) from the pure translational modes and highest IPR approaching $\sim$ 0.9 and with high frequency modes having IPRs exactly equal to 0.5. 
At $10^8$ K, nearly all modes (over 98\% of the mode populations) are `localized' in the conventional interpretation having IPR $>$ IPR\textsubscript{loc}.

\textit{Phase quotient}. -- Phase quotients for liquids and gases are shown in Fig. \ref{fig:PQ}B-D. For all systems, the phase quotient spectra appear nearly symmetric for real and imaginary modes with the exception of the long tail in the real frequency spectrum. A monotonic decrease in PQs is apparent moving away from 0 THz, similar to the glass system. One notable trend observed for liquids and gases vs. solids is the increasing presence of modes with PQ = -1 with increase in temperature. In both metrics (IPR and PQ), we do not observe distinct differences between real mode and imaginary mode characters. 

\begin{figure}[h!]
 	\centering
 	\includegraphics[width=1\linewidth]{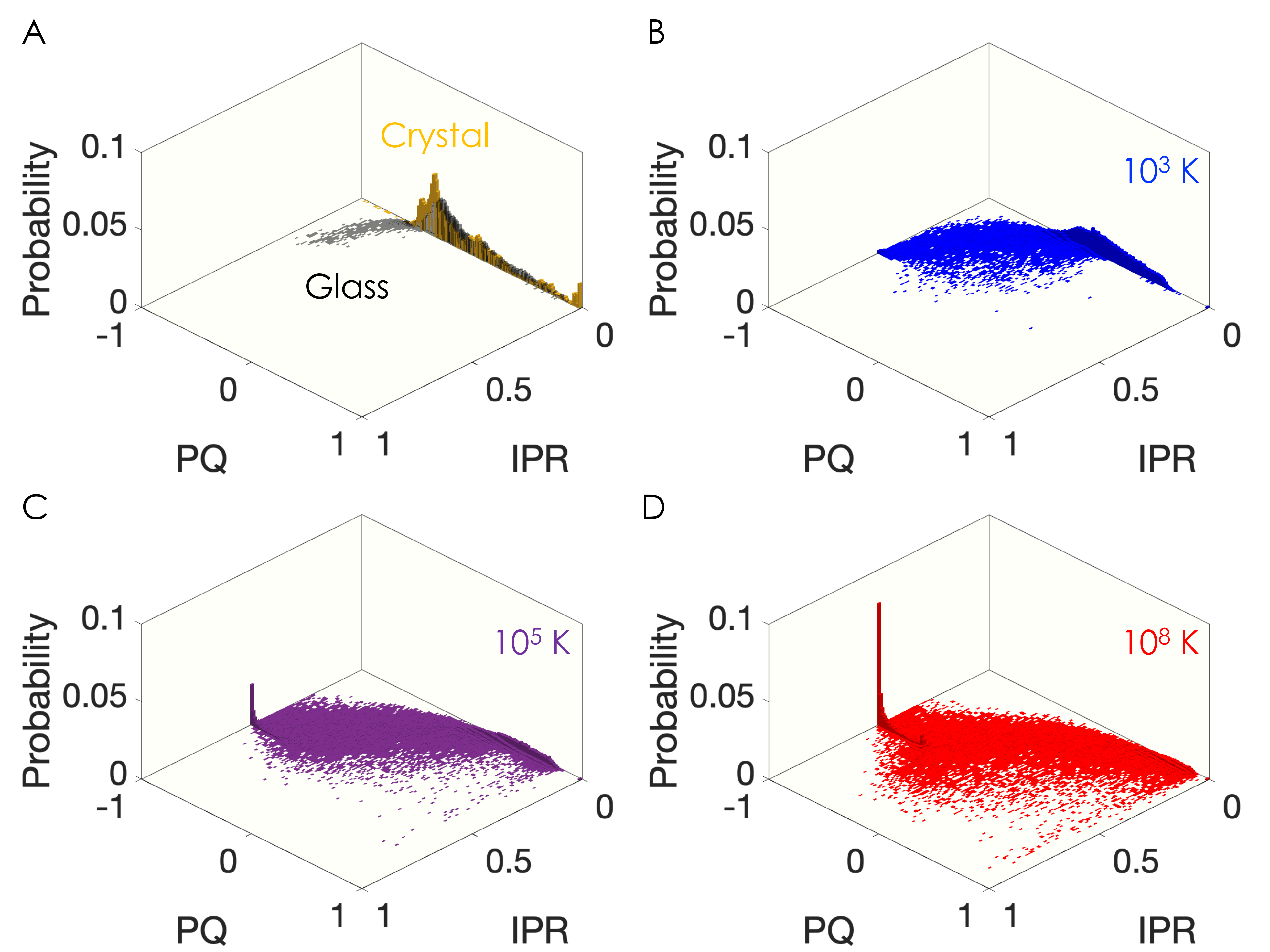}
 	\caption{Bivariate (inverse participation ratio and phase quotient) probability distributions (A) 1 K solids (yellow for crystal and black for glass) and (B-D) select high temperature liquids and gases at 10\textsuperscript{3} K (blue), 10\textsuperscript{5} K (purple), and  10\textsuperscript{8} K (red). We observe a peak appearing at IPR = 0.5 and PQ = -1 with increase in temperature.}
 	\label{fig:IPR_PQ}
 \end{figure}

Examining the gas modes more closely, we find some modes with exact values of IPR = 0.5 and/or PQ = -1, modes that are absent in the solid systems. Bivariate probability distributions of the IPRs and PQs for argon at select temperatures are plotted in Fig. \ref{fig:IPR_PQ} and demonstrate a strong trend: going from solids to gases, modes with both IPR = 0.5 and PQ = -1 become the dominant population. In the limit that $T \to \infty$, we imagine that this trend continues as atoms become effectively smaller and this character will govern the normal mode spectra. As our interest lies in using normal mode analysis to understand liquids in terms of higher T gas phases, we aim to fully understand these modes. 

\begin{figure}[h!]
 	\centering
 	\includegraphics[width=0.7\linewidth]{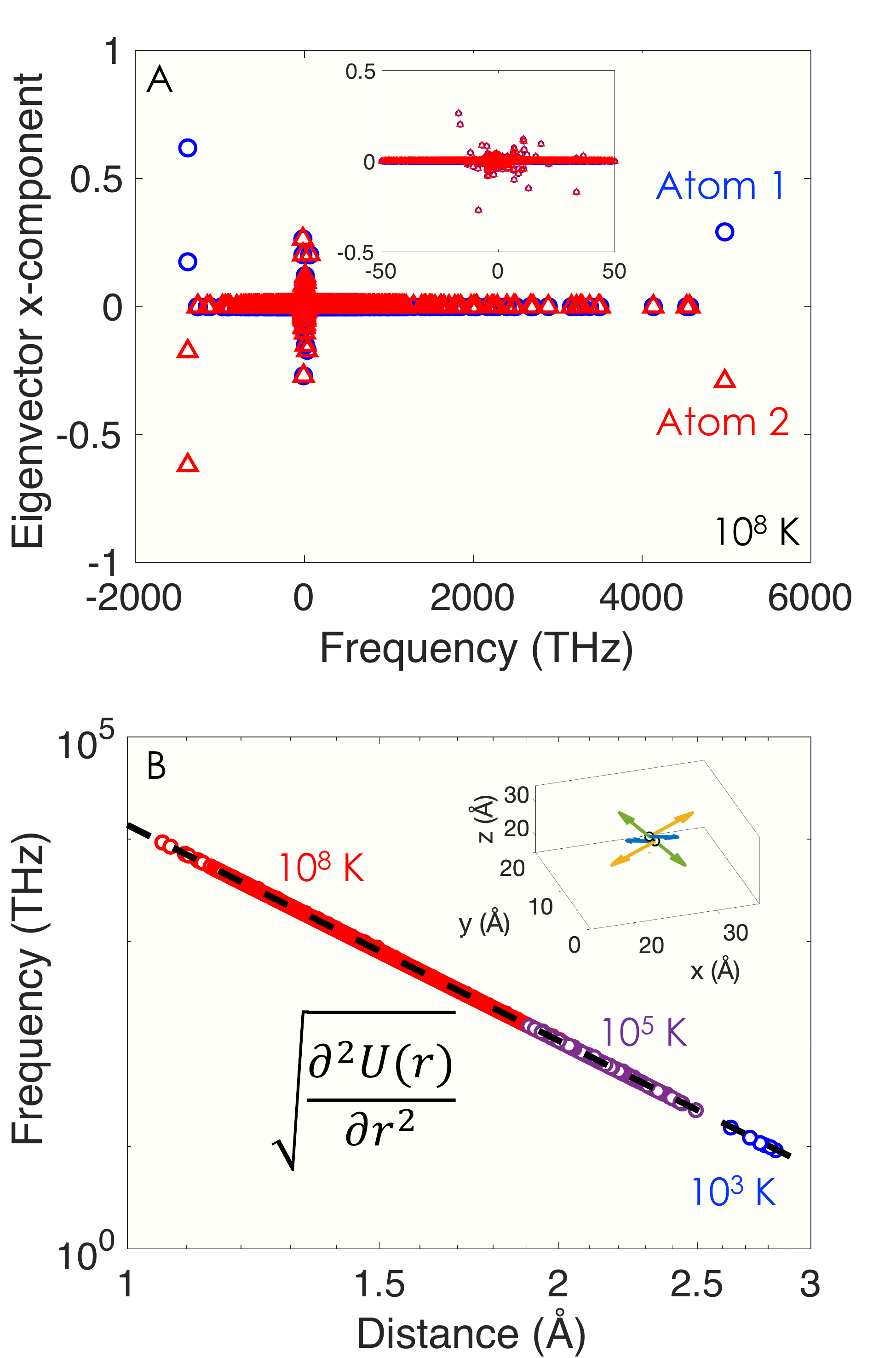}
 	\caption{(A) Eigenvector components ($x$-direction for a pair of atoms identified to have IPR = 0.5 and PQ = -1 ) for all modes at 10\textsuperscript{8} K. Blue and red colors represent atom 1 and atom 2 in the pair. Inset shows a zoomed-in view near 0 THz. All eigenvectors are in-phase between the two atoms except three which are fully out-of-phase. All pairs of atoms with these features are identified and their along-the-bond mode (green arrows in inset) frequencies and their distances are shown in (B). Clear relations to the LJ pair potential (black dashed line) is demonstrated.}
 	\label{fig:3modes}
 \end{figure}

\textit{IPR = 0.5 \& PQ = -1 modes}. -- IPR of 0.5 indicates that two atoms are participating in the mode and PQ = -1 describes eigenvectors among neighboring atoms that are fully out-of-phase.  From the simulations we have identified pairs of atoms that govern modes with IPR = 0.5 and PQ = -1 and examined the roles of these atoms in all modes of the system. As a representative example, Fig. \ref{fig:3modes}A gives ($x$-direction) eigenvector components for such a pair (atom 1: blue and atom 2: red) for all modes at $10^8$ K. This pair of atoms contributes negligibly to most modes (near zero amplitudes) as nearly all modes are localized to other atoms (large IPR values). However, for modes with finite and sometimes large $x$-components (see inset) for this pair, the eigenmotion of the two atoms are the same (overlapping components), while three modes have identical amplitudes in opposite direction. Thus, of the multitudes of other mode motions, their relative motions are defined by three modes with opposite directions (IPR = 0.5 and PQ = -1). $y$ and $z$ eigenvector components have similar features (not shown to avoid redundancy). 

Interestingly, the three modes that govern the motion of the pairs of atoms with IPR = 0.5 and PQ = -1 can be decomposed into one real mode and two imaginary modes. Visualization of these three modes projected on a pair of atoms is representatively shown in the inset of Fig. \ref{fig:3modes}B. The real modes have eigenvectors along the ‘bond’ (green arrows), while atomic motion for the other two imaginary modes are orthogonal to the bond. In essence, the repulsive force along the bond directs the atoms to a more stable position while there is no energy loss in the motion of the other two modes, similar to rotational modes. 

In bulk dense solids, each atom sits in an effective potential that is the summation of all neighboring interatomic potential interactions. However, for the pairs of atoms dominating the eigenmotions for IPR = 0.5 and PQ = -1 modes, the behavior is simulating independent two-atom motion governed by a single pair potential ($U_{i,j}(r)$) that defines the mode frequencies, while other interactions are negligible. This leads to three Goldstone modes (two atoms moving in-phase for three Cartesian degrees of freedom) and three out-of-phase modes with opposite eigenvector directions with one being along the bond. Mode frequency along the bond is strictly determined by $\frac{\partial^2 U_{i,j}(r)}{\partial r^2}$. Along-the-bond mode frequencies with IPR = 0.5 and PQ = -1 as a function of distance between the atomic pairs are shown in Fig. \ref{fig:3modes}B for both liquid and gas systems. The nearly perfect power law behavior demonstrates that these mode frequencies are purely determined by the second derivative of the pair potential, similar to truly independent two-atom interactions. We also observe that there is a temperature dependence in the bond distance and mode frequencies. There is a large distribution of bond lengths and frequencies at high temperature while the distribution is smaller for low temperatures. While dynamical matrices are built based purely on the potential of a static structure, kinetic energies play a dominant role in determining the atomic locations that feed into the dynamical matrices, especially at high temperatures (liquids and gases). The larger the kinetic energies that atoms have, the closer they can approach each other, leading to higher frequencies for these modes. 

\textit{Collisons}. -- Normal modes are typically used to describe phonon quasi-particles or harmonic oscillators in solids. Even in gases, there can be internal degrees of freedom for non-monoatomic species that lead to vibrations (e.g., H\textsubscript{2} and O\textsubscript{2}). However, the results described above show that along-the-bond modes with IPR = 0.5 and PQ = -1 should be considered as collisional rather than vibrational. Even with absence of kinetic energy, if pairs of argon atoms are initially placed at distances shown in Fig. \ref{fig:3modes}B and their motions are allowed to evolve with time, their potential energy is positive ($U_{LJ} (r < 3.4$ \AA $) > 0)$ such that they dissociate after participating in the along-the-bond mode motion with kinetic energies determined by energy conservation. We, thereby, characterize these modes as two-body collisional modes or `collisons'. 

\begin{figure}[h!]
 	\centering
 	\includegraphics[width=1\linewidth]{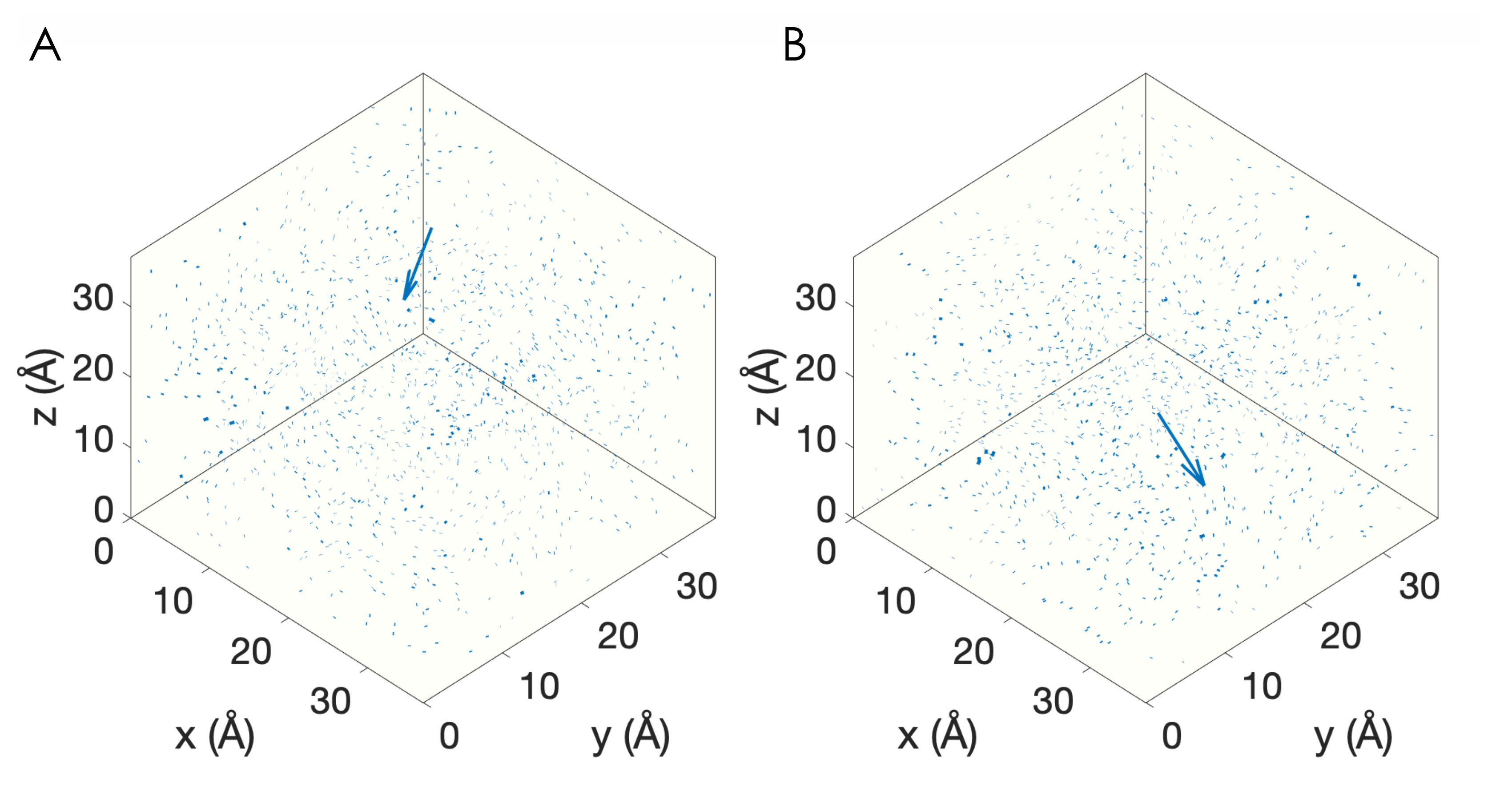}
 	\caption{Eigenvector maps of all atoms for (A) -0.3 THz mode with IPR = 0.82 and (B) 0.1 THz mode with IPR = 0.74, among the highest IPRs for $10^8$ K systems. Eigenvector amplitudes are multiplied by a factor of 10 for easier visualization, same as Fig. \ref{fig:3modes}B. We observe that for these modes, eigenvector magnitudes are concentrated to a single atom.}
 	\label{fig:HighIPR}
 \end{figure}

 \textit{Translatons.} -- As evident from Fig. \ref{fig:IPR}, the highest IPR values which are located near 0 THz also depend on temperature. Traditional interpretation of IPRs that IPR is a measure of the number of atoms participating in the mode does not strictly work here as there are many IPRs between $\frac{1}{2}$ (2 atoms participating) and 1 (1 atom participating). However, we find that these high IPR modes appear to have concentrated eigenvector amplitude on a single atom as demonstrated in Fig. \ref{fig:HighIPR}. 
 
It is, therefore, expected that under different conditions there exist modes with IPR approaching 1 where mode motion is projected to a single atom translating through space with small frequency. This is consistent with what we know about dilute gas eigenmodes where nearly all modes have zero eigenfrequencies due to lack of potential interactions. As we progressively reach lower densities, atomic interactions will become less frequent and more uninterrupted translational motion will appear, which would have $\sim$ zero eigenfrequency. In the same token, in the limit of $T \to \infty$ where atoms become point particle-like and overall collisions become less frequent, we would expect modes that purely describe translational motion, or what we may characterize as `translatons', to appear. We have done 10\textsuperscript{12} K IPR calculations (not shown) and found that IPR approaches as high as 0.97 vs. 0.85 observed in our 10\textsuperscript{8} K system. 

\section{Concluding remarks}

Our results on the normal mode analysis on heat capacities \cite{moon_microscopic_2022}, pair distribution functions, and mode characters by IPRs and PQs, suggest that in the limit of $T \to \infty$, particles will effectively behave as point-particles and two-body collisional modes dominate the particle dynamics, even when the volume is kept constant at that of a solid. This is consistent with real gas collisions where two-atom collisions become statistically dominant with increase in temperature over many-body collisions \cite{bonasera_competition_1991}. In glasses, eigenmodes are vibrational within a local harmonic well and have been categorized into propagons, diffusons, and locons depending on the degree of localization \cite{allen_thermal_1989, allen_diffusons_1999}. Although there exist different ways to categorize these modes \cite{seyf_method_2016, moon_propagating_2018, moon_examining_2021}, they often rely on the normal mode analysis, i.e., eigenvector characters. Similarly, we categorize eigenmodes at $T \to \infty$ into collisons and translatons from their eigenvector characters and their local atomic environment.

\begin{figure}[h!]
 	\centering
 	\includegraphics[width=1\linewidth]{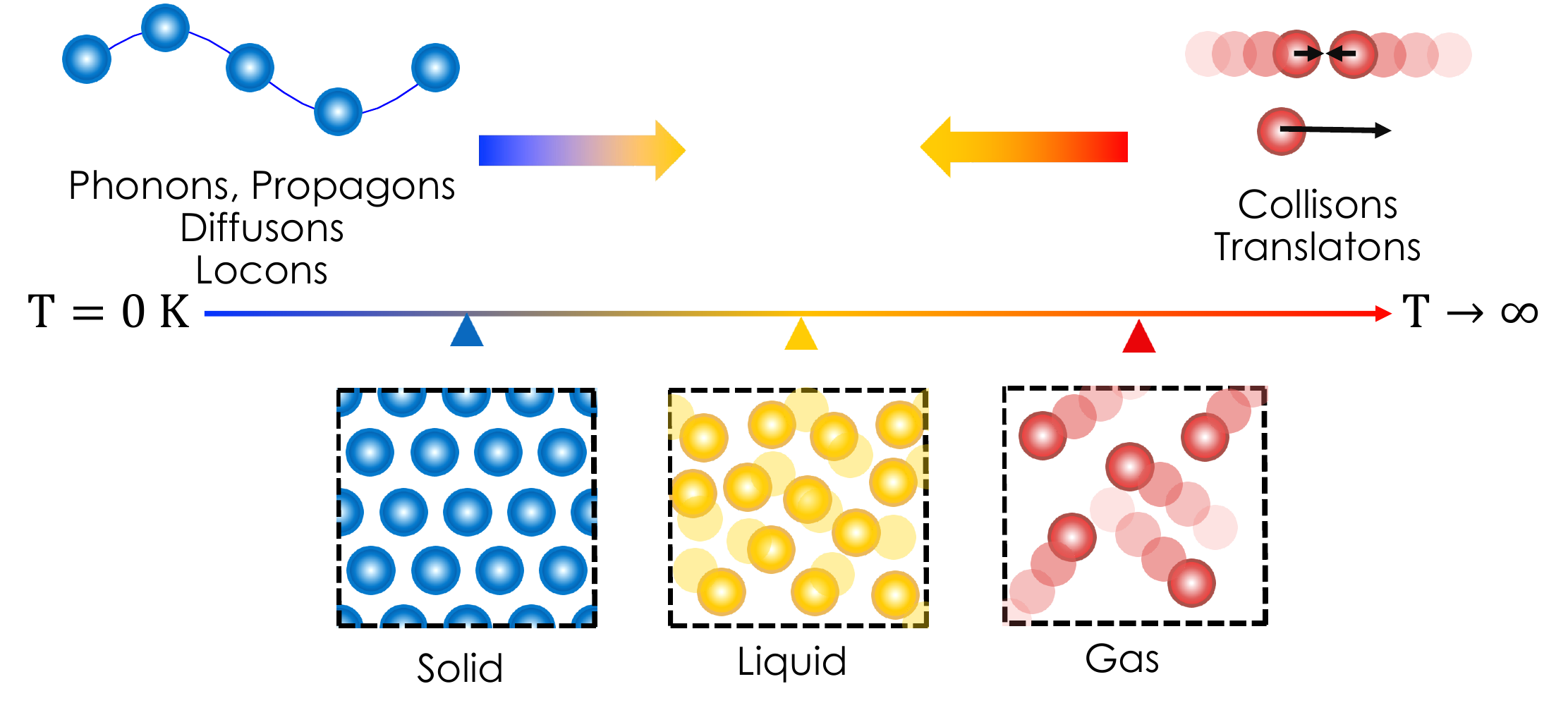}
 	\caption{Spectrum of eigenmodes of dynamical matrices from $T = 0$ K to $T \to \infty$ under constant volume. At $T = 0$ limit, we have vibrational modes (phonons, propagons, diffusons, and locons depending on disorder and localization). At the high temperature limit, both real and imaginary modes are expected to be gas-like, which we characterize as collisons and translatons. Eigenmodes for finite, intermediate temperature structures could potentially be described through these two limits.}
 	\label{fig:overview}
 \end{figure}

With these insights into the behavior of high temperature (gas) eigenmodes of dynamical matrices coupled with the vast body of research regarding T = 0 eigenmodes (solid), we propose that eigenmodes in liquids may be best considered as a collection of both solid-like vibrational and gas-like collisional and translational modes as summarized in Fig. \ref{fig:overview}. While the probability of having two-body collisional modes decreases with decreasing temperature, we still observe these modes even in the lowest temperature liquid argon considered here. In real liquid systems at lower densities, where potential interactions are weaker, we expect that normal mode spectra will still consist of some gas-like modes. In this work, a simple Lennard-Jones pair potential was used throughout. We anticipate that our high temperature limit results will not depend on the choice of interatomic potential (two-body, many-body, etc) as all systems will become an atomic gas where details of the nature of the interactions become less important. 

In this work, we explored the high temperature limits of eigenmodes of dynamical matrices through the lens of inverse participation ratios and phase quotients in the context of $T \sim 0$ K eigenmodes. In crystals and glasses, normal modes are considered to be vibrational in nature with most being delocalized in space. With increase in temperature we find that an increasing number of eigenmodes become gas-like collisional and translational modes. Our work, therefore, sets the scale of spectra of different eigenmodes such that normal modes of liquids can be described in terms of evolving characters, rather than extrapolation from our understanding of normal modes in solids alone as often done in literature.

\section*{ACKNOWLEDGEMENT}
Authors thank Simon Th\'ebaud for helpful discussions on the nature of normal modes. This research was supported by the U.S. Department of Energy, Office of Science, Basic Energy Sciences, Materials Sciences and Engineering Division. This work used the Extreme Science and Engineering Discovery Environment (XSEDE) Expanse under Allocation No. TG-MAT200012. This research used resources of the National Energy Research Scientific Computing Center (NERSC), a U.S. Department of Energy Office of Science User Facility located at Lawrence Berkeley National Laboratory, operated under Contract No. DE-AC02-05CH11231 using NERSC award BES-ERCAPERCAP0020503.

\section*{Data Availability}
All data are included in the text.
\clearpage

\clearpage

\clearpage

\clearpage

\clearpage

\clearpage

\clearpage

\bibliography{apssamp.bib}

\end{document}


\title{Supplementary Materials for\\
Microscopic view of heat capacity of matter: solid, liquid, and gas}

\author{Jaeyun Moon}
\email{To whom correspondence should be addressed; E-mail:  moonj@ornl.gov}
 \affiliation{Materials Science and Technology Division,\\ Oak Ridge National Laboratory, Oak Ridge, Tennessee 37831, USA}
\author{Simon Th\'ebaud}
 \affiliation{Materials Science and Technology Division,\\ Oak Ridge National Laboratory, Oak Ridge, Tennessee 37831, USA}
\author{Lucas Lindsay}
 \affiliation{Materials Science and Technology Division,\\ Oak Ridge National Laboratory, Oak Ridge, Tennessee 37831, USA}
 
\author{Takeshi Egami}
\affiliation{Materials Science and Technology Division,\\ Oak Ridge National Laboratory, Oak Ridge, Tennessee 37831, USA\\
 Department of Materials Science and Engineering,\\ University of Tennessee, Knoxville, Tennessee 37996, USA\\
 Department of Physics and Astronomy,\\ University of Tennessee, Knoxville, Tennessee 37996, USA}

\date{\today}

\maketitle
\clearpage

\renewcommand{\figurename}{Fig.}
\renewcommand{\thefigure}{S\arabic{figure}}
\setcounter{figure}{0}

\section{Simulation details}

For argon, FCC structure of 1372 atoms with a lattice parameter of 5.2686 \AA \, ($\rho_0$) was used as an initial input structure for molecular dynamics simulations. Temperatures considered were from 1 to $10^8$ K in increments of factors of 10 (log scale) excluding 100 K near the melting temperature. Extremely high temperatures were necessary to reach the gas limit ($C_V = 1.5Nk_B$). MD Timesteps were 0.1 and 0.01 fs depending on the temperature to capture the fast atomic dynamics. For silicon, FCC structure with 1728 atoms with lattice parameter of 5.431 \AA \, ($\rho_0$) was employed. Calculations were done at temperatures from 1 to $10^7$ K with varying timesteps of 0.5 and 0.05 fs. For iron, BCC structure with 2000 atoms with a lattice parameter of 2.867 \AA \, ($\rho_0$) was studied from 1 to $10^6$ K. Timesteps of 0.1 and 0.01 fs were used.  At each temperature, all systems were equilibrated for $5 \times 10^6$ timesteps in the canonical ensemble (NVT) prior to data recording of another $5 \times 10^6$ timesteps under the same ensemble. Relativistic corrections were not needed at high temperatures due to relatively heavy masses of our systems. 

These systems have different atomic structures in the solid state: face centered cubic (argon), face centered cubic (silicon), and body centered cubic structures (iron) below melting temperatures under atmospheric pressures. Upon melting, liquid argon and iron atoms are close packed with coordination numbers around 13 to 14. In comparison, silicon possesses more complex temperature dependent structural features. Due to formation of metallic bonds upon melting, the coordination number of silicon increases from 4 to 6 or more and there also exists a low density liquid to high density liquid transition, commonly observed in other tetrahedral systems including water. Both structural changes in silicon have been well-characterized by the Stillinger-Weber potential used here. Having the various structural features and bond natures described above, argon, silicon, and iron are versatile test systems for the purpose of this study.

\section{Diffusion coefficients and instability factors}

Temperature dependent self-diffusion coefficients, $D$, for all $\rho_0$ systems were calculated by 
\begin{equation}
    D = \lim_{t \to \infty}\frac{1}{6t}\langle [\boldsymbol{r}_i(t) -  \boldsymbol{r}_i(0)]^2 \rangle 
\end{equation}
where $t$ is time and $\boldsymbol{r}_i(t)$ is the time dependent position of atom $i$. As shown in Fig. \ref{fig:Diffusion}, we observe a continuous increase in $D$ for all systems with increase in temperature. Error bars are smaller than the symbol sizes.

\begin{figure}
	\centering
	\includegraphics[width=0.7\linewidth]{Diffusion.pdf}
	\caption{Temperature dependent self-diffusion coefficients, $D$, for all $\rho_0$ systems. Yellow circles, blue squares, and pink crosses are for argon, silicon, and iron, respectively. Monotonic increase in $D$ with increase in temperature is observed.}
	\label{fig:Diffusion}
\end{figure}

Prior works have attempted to relate self-diffusion coefficients, $D$, of liquids to the fraction of imaginary modes over the total number of modes, $3N$ \cite{seeley_isobaric_1991,nave_instantaneous_2000,clapa_localization_2012}. Some have argued that $D$ is directly proportional to this fraction \cite{seeley_isobaric_1991} while others have argued for more subtle relations between $D$ and the fraction of `delocalized' imaginary modes \cite{clapa_localization_2012}. However, INM($\omega$) for our high temperature liquid and gas systems challenge these ideas: (1) we see a dramatic slow-down of how fast the density of states changes with temperature as evident in Fig. 2 (A) in the main text where temperature increases evenly by a factor of 10 but self-diffusion coefficients at these temperatures continue to increase with temperature as shown in Fig. S1. (2) All imaginary modes in gas states at high temperatures are localized as confirmed by their inverse participation ratios (See Section V). 

More recently, Zaccone and Baggioli reported a linear power law at low frequencies in INM($\omega$) \cite{zaccone_universal_2021} and suggested that imaginary modes are important in describing heat capacity in liquids \cite{baggioli_explaining_2021}. However, their work relies on empirical model fitting, leading to difficulties in understanding the nature of heat carriers in liquids. Schirmacher et al. emphasized the importance of relative shapes of real and negative eigenvalue distributions of the dynamical matrices in understanding the nature of vibrational modes in liquids and how they become more symmetric with increase in temperature in liquids above glass transition temperatures \cite{schirmacher_modeling_2022}.

Rather than characterizing specific shapes of certain instantaneous normal mode spectra, we focus on understanding the demographics of total number of atomic degrees of freedom in liquids: solid-like or gas-like. We have introduced two parameters describing gasness of a system designated as instability factors given by $\text{IF}_1 = \frac{2N_i}{3N}$ and $\text{IF}_2 = \frac{N_i}{3N-N_i}$, where $N_i$ is the total number of imaginary modes. Temperature dependence of these parameters are shown in Fig. \ref{fig:IP}. We observe a clear slowdown in increase in both instability factors with temperature around $\text{IF}_{1,2} = 1$. Instability factors are slightly above 1 at high temperatures. We believe that this is due to small statistics arising from using relatively small system sizes. To adequately test this issue, we would need to do lattice dynamics calculations on much bigger systems of $N = 10^4$ to $N = 10^6$ atoms; however, diagonalizing such large $3N$ by $3N$ matrices is challenging. We have additionally done lattice dynamics calculations of argon gas at a much lower density of $0.1 \rho_0$ from 500 K to $10^4$ K and $\text{IF}_{1,2}$ were constant at $\sim 1.1$.

\begin{figure}[h!]
	\centering
	\includegraphics[width=0.7\linewidth]{IF.pdf}
	\caption{Temperature dependent instability factors for all $\rho_0$ systems. Yellow circles, blue squares, and pink crosses are for argon, silicon, and iron, respectively.}
	\label{fig:IF}
\end{figure}

\clearpage

\section{INM($\omega$) and VACF($\omega$) equivalence for solids}

Instantaneous normal mode and velocity autocorrelation spectra for solids are known to be equivalent under harmonic approximations and are both used to obtain phonon density of states of solids. An example of this equivalence is shown in Fig. \ref{fig:DOS}(A) for crystalline silicon at 1 K with the Stillinger-Weber potential treating the entire domain as a unit cell, and considering the Gamma point only. On the other hand, strikingly different spectra are observed for liquid and gas systems as demonstrated in Fig. \ref{fig:DOS}(B). In this work, we make connections between these seemingly different spectra in liquids and gases via instability factors characterizing the nature of the atomic degrees of freedom. 

\clearpage 

\begin{figure}[h!]
	\centering
	\includegraphics[width=1\linewidth]{INM_VACF.pdf}
	\caption{(A) Phonon density of states for crystalline silicon at 1 K. We see nearly identical spectra between INM($\omega$) and VACF($\omega$), as expected. (B) INM($\omega$) and VACF($\omega$) of silicon gas at $10^7$ K. The Stillinger-Weber potential was used. Negative frequencies denote modes with imaginary frequencies.}
	\label{fig:DOS}
\end{figure}

\clearpage
\section{Temperature dependent specific heat}

For canonical ensembles in equilibrium, statistical mechanics dictates that specific heat is related to energy fluctuations by 
\begin{equation}
    C_V = \frac{\langle E^2 \rangle - \langle E \rangle^2 }{k_BT^2}
\end{equation}
where angled brackets represent ensemble averages. Temperature dependent specific heats of all systems studied in this work are shown in Fig. \ref{fig:Cv_T}. Melting temperatures are denoted by the red lines on the $x$-axis. For molecular dynamics simulations, which are classical, specific heats for solids even at low temperatures are $3Nk_B$ as observed in the figure. We observe clear transitions to the gas limit as we increase the temperature. 

Instability factors versus constant volume specific heats are shown in Fig. \ref{fig:IF_Cv}. Despite different spectral shapes and features among argon, silicon, and iron systems as demonstrated in Fig. 2 and Fig. 3, strong correlations between instability factors and heat capacities with small spread are evident. As predicted, specific heat goes to the gas limit of $1.5Nk_B$ near IF\textsubscript{1,2} = 1.

\begin{figure}
	\centering
	\includegraphics[width=0.5\linewidth]{Cv_T.pdf}
	\caption{Temperature dependent constant volume heat capacity of (A) argon, (B) silicon, and (C) iron with different densities. Different marker colors represent different densities. Brown dash lines and black dash-dot lines represent Dulong-Petit limits and gas-limits for monatomic systems. $T_{c \: \rightarrow \: l}$ marked by red lines mean crystal to liquid melting temperatures.}
	\label{fig:Cv_T}
\end{figure}

\begin{figure}[h!]
	\centering
	\includegraphics[width=0.7\linewidth]{IF_Cv.pdf}
	\caption{Instability factors versus constant volume specific heats.}
	\label{fig:IF_Cv}
\end{figure}

\clearpage

\section{Instantaneous normal mode characters from solid to gas}
In this section, we provide some insights into the instantaneous normal mode character via some widely used measures including inverse participation ratios and phase quotients. Inverse participation ratio (IPR) is a measure of how many atoms participate in the motion of a particular mode. IPR  has been used to qualitatively provide information about a degree of localization of a mode in various complex systems \cite{moon_sub-amorphous_2016, deangelis_thermal_2018}. IPRs ($p_n^{-1}$) are found by \cite{bell_atomic_1970}
\begin{equation}
    p_n^{-1} = \sum_i \Big( \sum_\alpha e_{i\alpha, n}^* e_{i\alpha, n} \Big)^2
    \label{Eq: IPR}
\end{equation}
where $e_{i\alpha, n}$ is the eigenvector component for atom $i$ in $\alpha$ direction for the mode $n$. The IPR is conventionally interpreted that it is equivalent to $1/N$ if all atoms are participating in the mode with $N$ being the number of atoms, or 1 if the mode motion is completely localized to one atom. In literature, a normal mode has been considered to be localized when less than 10 to 20\% of atoms are participating in the mode motion \cite{allen_diffusons_1999, seyf_method_2016, aryana_tuning_2021}. As IPR analysis is rarely used for crystalline materials, we instead first show IPR values of our $\rho_0$ glass systems at 1 K in Fig. \ref{fig:IPR_glass} for more meaningful comparisons against other phases later. For our structures, IPR values over IPR\textsubscript{loc} = 0.007, 0.006, and 0.005 corresponding to 10\% of total number of atoms participating are considered to be localized for argon, silicon, and iron, respectively. All glass systems appear to have similar spectral trends. At 0 THz, we have the Goldstone modes where all atoms are participating in the purely translational motion, leading to IPR = $1/N$. Most low frequency modes are spatially extended with a few exceptions and have IPR values above IPR\textsubscript{loc}. These exceptions are also known as quasi-localized or resonant modes. Some prior works have proposed that some localization at low frequencies is due to a finite size effect \cite{feldman_numerical_1999}. At high frequencies, we observe a large increase in IPR values over IPR\textsubscript{loc}. These localized modes, often referred to as locons, in glass literature form a small fraction of the total number of modes about 2 to 3\% in these systems, consistent with prior works \cite{allen_diffusons_1999, moon_examining_2021, lv_direct_2016, larkin_thermal_2014}. Locons typically have low mode thermal diffusivity and conductivity compared to extended modes \cite{allen_diffusons_1999, moon_propagating_2018, kim_origin_2020}. Maximum IPR values found in our glasses are around 0.3.  

With the IPR trends for the glasses as the background information, IPR spectra at various temperatures for liquid and gas systems are shown in Fig. \ref{fig:IPR}. The selected temperatures are the same as those shown in Fig. 3 in the main text. At the lowest temperatures shown, we see asymmetric U shapes in the IPR spectra. As we approach low magnitude in frequency close to zero, modes appear to be more extended. Similar trends have been previously shown in other liquid systems \cite{zhang_what_2019}. In the limit of high frequency magnitude, we seem to approach IPR $\sim$ 0.5. With increase in temperature, low frequency magnitude modes have higher IPR values, filling up the empty region in the U shapes found in glass systems. At the highest temperatures, nearly all modes (over 98\% of the mode populations) are `localized' in the conventional interpretation having IPR $>$ IPR\textsubscript{loc}. If IPR\textsubscript{loc} corresponding to 20\% of atoms participating in the mode as was done before in literature \cite{aryana_tuning_2021}, all our modes would be considered localized except the three translational modes for our highest temperature systems. Interestingly, asymptotic limit of IPR values at high frequency magnitudes becomes very clear having the exact value of IPR = 0.5. At lower frequencies, some modes appear to have IPR $>$ 0.5. 

To gain more insights into the overall changes in the IPR spectra across different phases, we show the probability distributions of IPR values at select temperatures (10\textsuperscript{3}, 10\textsuperscript{5}, and 10\textsuperscript{8} K) for argon as a representative case. For the lowest temperature, it is clear that close to zero IPR values dominate. A similar trend is observed for glass systems (not shown). Once the temperature is increased, however, we start to see another peak appearing at IPR = 0.5 while the original close-to-zero peak significantly reduces. There are also some overall increases in probability density functions at intermediate IPRs. Since they appear to be the most noticeable changes in the IPR spectra from glasses and low temperature liquids to high temperature gases, we focus our subsequent discussions on characterizing IPR = 0.5 modes. 

In addition to the inverse particiation ratios, another metric called phase quotient (PQ) has been useful in characterizing normal modes in disordered materials \cite{bell_acoustic_1975, allen_diffusons_1999, seyf_importance_2018} and is defined as \cite{bell_acoustic_1975}
\begin{equation}
    PQ_n = \frac{\sum_m \sum_\alpha e_{i\alpha, n} e_{j\alpha, n}}{\sum_m \Big|\sum_\alpha e_{i\alpha, n} e_{j\alpha, n}\Big|}
\end{equation}
where the eigenvector subscripts are identical to those in Eq. \ref{Eq: IPR} and the sum over $m$ represents counting only the nearest neighbor pairs among atoms $i$ and $j$. Prior works utilizing PQs in glasses have proposed that modes with PQ $>$ 0 and PQ $<$ 0 represent acoustic-like and optical-like modes, respectively \cite{allen_diffusons_1999, seyf_importance_2018}. More generally, PQ is a measure of the phase relationship of a mode eigenvector of an atom and its neighbors. PQ value of 1 then represents cluster of atoms all moving in-phase which is the case for Goldstone translational modes and PQ value of -1 means that all atoms participating in the mode are moving in the opposite phase/direction with their nearest neighbors. For our systems, we define the nearest neighbors when they are within $d_{c} = 2r_c$ where $\frac{4}{3}\pi r_c^3 = \frac{V}{N}$ but we find that our results are not very sensitive of the choice of $d_c$. Modal phase quotients for various argon systems are shown in Fig. \ref{fig:PQ}. For the argon glass, we observe a monotonic decrease in PQ values with increase in frequency, similar to other glasses reported in literature \cite{allen_diffusons_1999, seyf_importance_2018}. We have PQ = 1 for the Goldstone translational modes as expected but we do not seem to have any modes with PQ = -1. In crystalline and amorphous solids, modes with PQ = -1 are mostly non-existent (for example in crystalline silicon, amorphous carbon, amorphous silica, and crystalline InGaAs alloy) and only one a single mode has been shown to have PQ = -1 in crystalline InAs and GaAs \cite{allen_diffusons_1999, seyf_importance_2018, bodapati_vibrations_2006}. With increase in temperature, we start to observe a large number of modes with PQ = -1 in our argon systems. Imaginary and real modes do not appear to have distinguishable differences in PQ. 

Some noticeable changes we have observed so far in inverse participation ratios and phase quotients from glass to gas are that we have a large number of modes with IPR = 0.5 and PQ = -1. Perhaps, there are some correlations between these two features. Bivariate probability distributions of the inverse participation ratios and phase quotients for argon at select temperatures are, therefore, plotted in Fig. \ref{fig:IPR_PQ_hist}. A clear trend is seen: as we go from glass to gas, modes with both IPR = 0.5 and PQ = -1 become the dominant population. IPR of 0.5 conventionally means 2 atoms are participating in the mode and PQ = -1 means eigenvector among neighboring atoms is fully out-of-phase. With all the evidence we have: pair distribution functions, specific heats, inverse participation ratios, and phase quotients, a clear physical picture of these modes then emerges: modes with IPR = 0.5 and PQ = -1 in the high temperature systems depict atomic collision events involving two atoms. A visual confirmation is shown in Fig. \ref{fig:eig_map} where eigenvector (brown arrows) for a mode is projected onto all the atoms (black circles). It is worth mentioning that the black circle diameter does not physically represent the effective diameter of atoms. Black circles are for visual aids only. For glasses, eigenvector for a locon is mostly spatially localized but it still involves a number of atoms in the mode motion. In addition to atoms located near $(x,y,z) = (30, 30, 10)$ \AA, there are still finite eigenvector magnitudes for atoms located near $(10, 20, 20)$ \AA, though not clearly depicted in Fig. \ref{fig:eig_map} A. For modes with IPR = 0.5 and PQ = -1 in argon gas, exactly two neighboring atoms moving in opposite directions are clearly shown, confirming our physical picture that these modes represent atomic collisions. All other atoms have zero components in the mode eigenvector. 

Normal modes are typically used to describe phonon quasi-particles in literature and various instantaneous normal mode works in liquids typically adopt the harmonic oscillator concept in understanding what instantaneous normal modes represent \cite{baggioli_explaining_2021, zaccone_universal_2021, stassen_instantaneous_1994, melzer_instantaneous_2012}. Our work here explicitly demonstrates that normal modes can also describe atomic collision motion rather than only simple harmonic oscillations, extending the use of normal mode analysis in describing atomic motion beyond phonon quasi-particles. 

\clearpage
\begin{figure}[h!]
	\centering
	\includegraphics[width=0.95\linewidth]{IPR_glass.png}
	\caption{Inverse participation ratios of (A) argon, (B) silicon, and (C) iron glasses. All the structures are from 1 K with $\rho_0$.}
	\label{fig:IPR_glass}
\end{figure}

\begin{figure}[h!]
	\centering
	\includegraphics[width=0.95\linewidth]{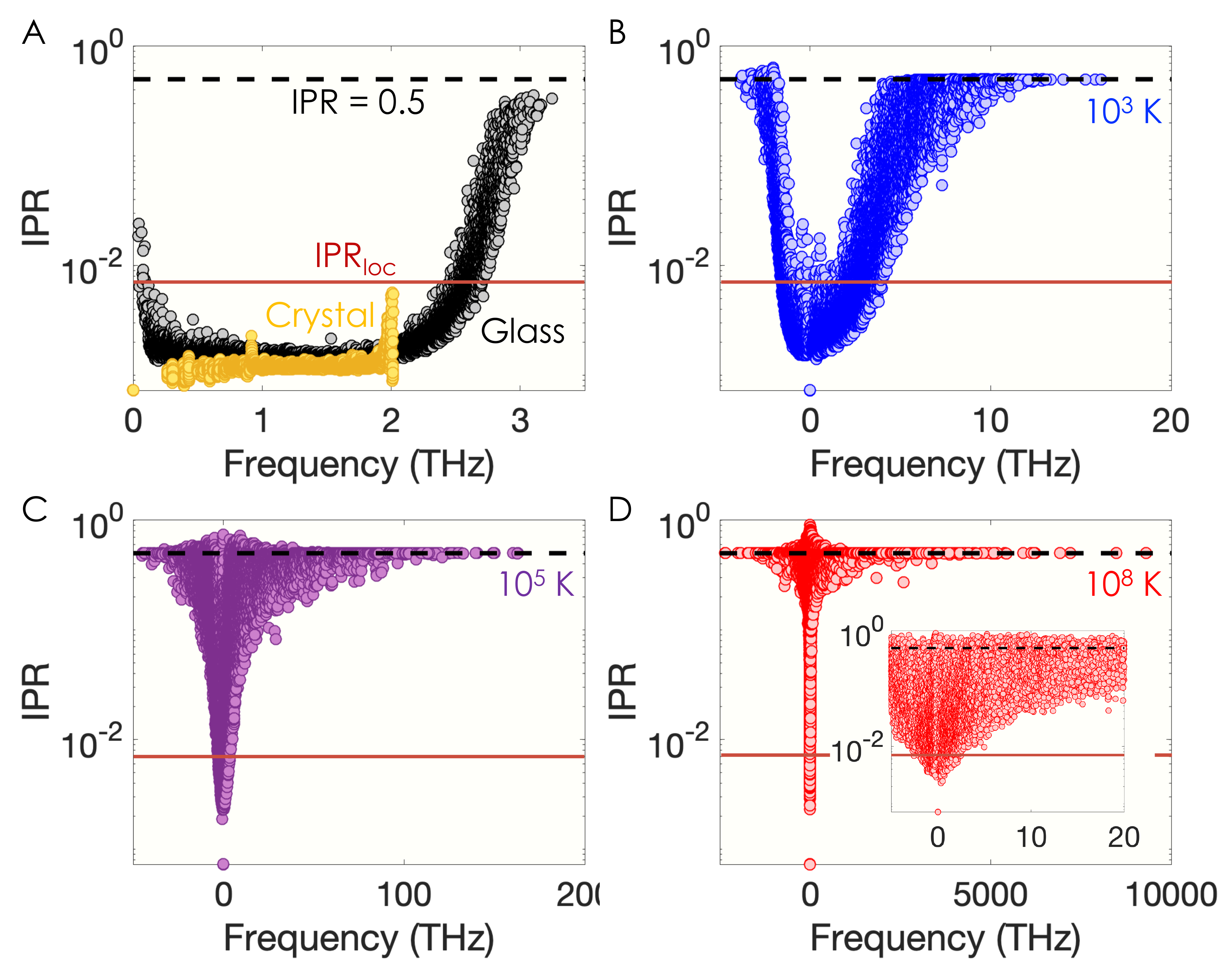}
	\caption{Inverse participation ratios at select temperatures of (A-C) argon at 10\textsuperscript{3}, 10\textsuperscript{5}, and 10\textsuperscript{8} K, (D-F) silicon at $5 \times 10^3$, $5 \times 10^4$, and 10\textsuperscript{7} K, and (G-I) iron at $5 \times 10^3$, $5 \times 10^4$, and 10\textsuperscript{6} K. All systems were evaluated at $\rho_0$. Inset figures are scaled such that horizontal and vertical axes are the same as those in the lowest temperature sub-figures for comparisons. Dashed lines indicate IPR = 0.5.}
	\label{fig:IPR}
\end{figure}

\clearpage
\begin{figure}[h!]
	\centering
	\includegraphics[width=0.5\linewidth]{IPR_hist.pdf}
	\caption{Inverse participation ratio probability distributions at select temperatures of (A-C) argon at 10\textsuperscript{3}, 10\textsuperscript{5}, and 10\textsuperscript{8} K. Arrows indicate the location of the maximum probability.}
	\label{fig:IPR_hist}
\end{figure}
\clearpage

\begin{figure}[h!]
	\centering
	\includegraphics[width=1\linewidth]{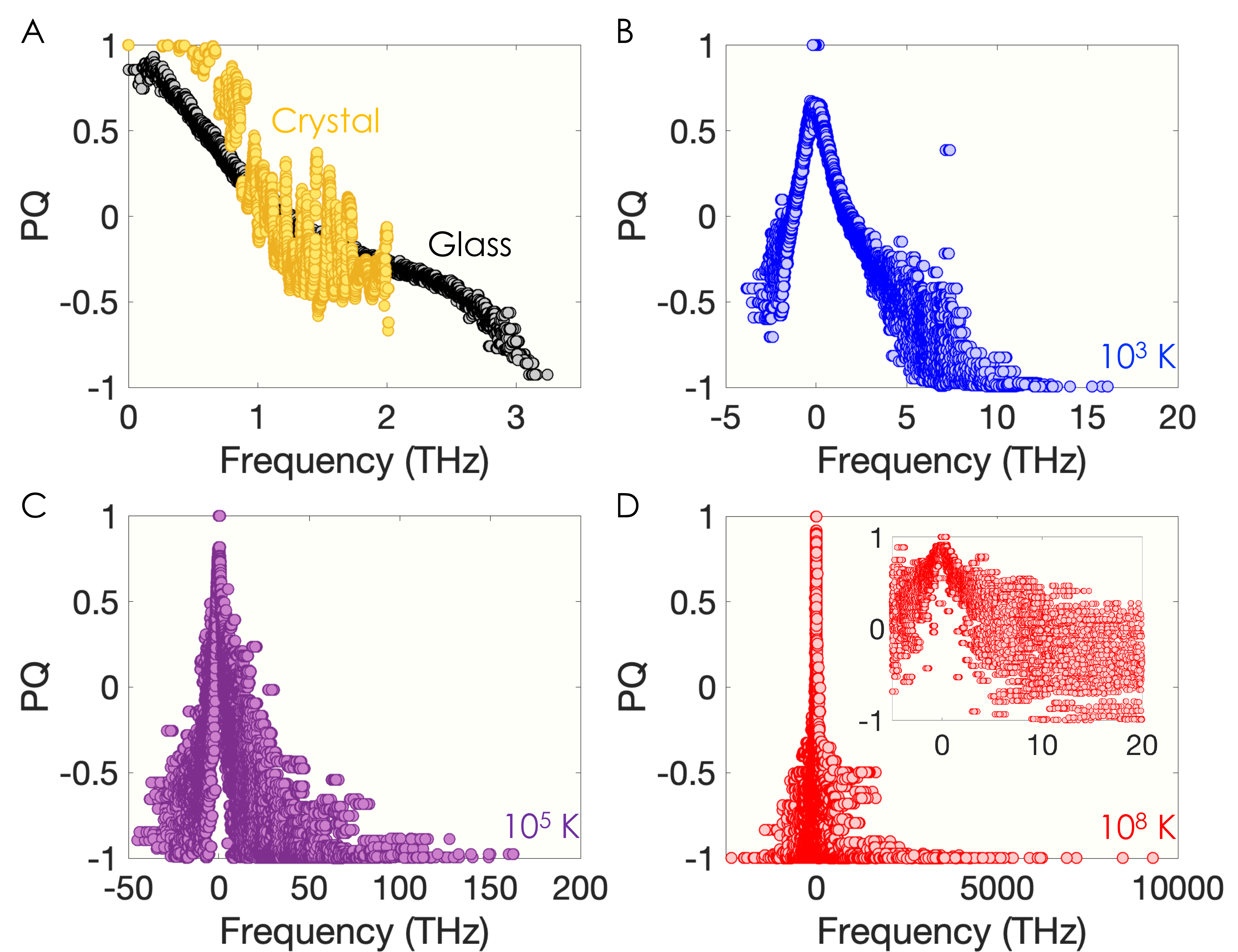}
	\caption{Phase quotients for various argon systems at (A) 1 K (glass), (B) 10\textsuperscript{3} K, (C) 10\textsuperscript{5} K, and (D) 10\textsuperscript{8} K.}
	\label{fig:PQ}
 \end{figure}
\clearpage

\begin{figure}[h!]
	\centering
	\includegraphics[width=1\linewidth]{IPR_PQ_hist.pdf}
	\caption{Bivariate (inverse participation ratio and phase quotient) probability distributions for various argon systems at (A) 1 K (glass), (B) 10\textsuperscript{3} K, (C) 10\textsuperscript{5} K, and (D) 10\textsuperscript{8} K. }
	\label{fig:IPR_PQ_hist}
 \end{figure}
\clearpage

\begin{figure}[h!]
	\centering
	\includegraphics[width=0.52\linewidth]{eig_map.pdf}
	\caption{Mode eigenvector for (A) 1 K argon glass at 3 THz with PQ = -0.8 and IPR = 0.3 which are among the highest IPR and lowest PQ values for the glass, (B) 10\textsuperscript{8} K argon gas at -69 THz, and (C) 10\textsuperscript{8} K argon gas at 65.8 THz. Both real and imaginary modes for the argon gas has IPR = 0.5 and PQ = -1. All the eigenvector components for all atoms are scaled by a factor of 10 for better visualization.}
	\label{fig:eig_map}
\end{figure}
\clearpage


\bibliography{apssamp.bib}